# Ultrasensitive Tunability of the Direct Bandgap of Two-dimensional InSe Flakes *via* Strain Engineering

Yang Li[1,§], Tianmeng Wang[1,§], Meng Wu[2,3§], Ting Cao[2,3], Yanwen Chen[1], Raman Sankar[4,5], Rajesh K. Ulaganathan[4,5], Fangcheng Chou[4,5], Christian Wetzel[6], Cheng-Yan Xu[7,*], Steven G. Louie[2,3,*], Sufei Shi[1,8,*]

[1.] Department of Chemical and Biological Engineering, Rensselaer Polytechnic Institute, Troy, NY 12180
[2.] Department of Physics, University of California at Berkeley, Berkeley, CA 94720
[3.] Materials Sciences Division, Lawrence Berkeley National Laboratory, Berkeley, CA 94720
[4.] Institute of Physics, Academia Sinica, Nankang, Taipei, Taiwan 11529
[5.] Center for Condensed Matter Science, National Taiwan University, Taipei, Taiwan 10617
[6.] Department of Physics and Astronomy, Rensselaer Polytechnic Institute, Troy, NY 12180
[7.] School of Materials Science and Engineering, Harbin Institute of Technology, Harbin 150001, China
[8.] Department of Electrical, Computer and System Engineering, Rensselaer Polytechnic Institute, Troy, NY 12180

[*]Corresponding Authors: shis2@rpi.edu, sglouie@berkeley.edu, cy_xu@hit.edu.cn

[§]These authors contributed equally to this work.

## Abstract

InSe, a member of the layered materials family, is a superior electronic and optical material which retains a direct bandgap feature from the bulk to atomically thin few-layers and high electronic mobility down to a single layer limit. We, for the first time, exploit strain to drastically modify the bandgap of two-dimensional (2D) InSe nanoflakes. We demonstrated that we could decrease the bandgap of a few-layer InSe flake by 160 meV through applying an in-plane uniaxial tensile strain to 1.06% and increase the bandgap by 79 meV through applying an in-plane uniaxial compressive strain to 0.62%, as evidenced by photoluminescence (PL) spectroscopy. The large reversible bandgap change of ~ 239 meV arises from a large bandgap change rate (bandgap strain coefficient) of few-layer InSe in response to strain, ~ 154 meV/% for uniaxial tensile strain and ~ 140 meV/% for uniaxial compressive strain, representing the most pronounced uniaxial strain-induced bandgap strain coefficient experimentally reported in two-dimensional materials.We developed a theoretical understanding of the strain-induced bandgap change through first-principles DFT and GW calculations. We also confirmed the bandgap change by photoconductivity measurements using excitation light with different photon energies. The highly tunable bandgap of InSe in the infrared regime should enable a wide range of applications, including electro-mechanical, piezoelectric and optoelectronic devices.

1. Introduction

The success of graphene research has inspired tremendous interest in two-dimensional (2D) materials beyond graphene, particularly 2D semiconductors with finite-size bandgap and high carrier mobility[1-4]. One particularly influential effort is on the transition metal dichalcogenides (TMDCs), which exhibit a transition from an indirect bandgap to a direct bandgap semiconductor when the TMDC flake is thinned down to the monolayer limit[1,4,5]. Monolayer TMDCs possess a unique valley-spin degree of freedom[5,6] and strong light-matter interaction[7], which can be exploited for novel quantum optoelectronic applications. However, having the direct bandgap in a limited visible spectrum range and limited carrier mobility compromise the prospect of utilizing TMDCs for light emitting device, solar energy harvesting or quantum electronic device applications[1,8,9]. In parallel, thin black phosphorous (BP) flake has been re-discovered as a new 2D material with carrier mobility as high as 6,000 $cm^2/V\cdot s$ which enabled the observation of quantum interference effect and quantum Hall effect (QHE)[3]. Optically, BP remains a direct bandgap material with an optical bandgap sensitively dependent on layer numbers, approaching the bulk value of ~ 0.3 eV from the monolayer value of ~ 1.5 eV quickly as the layer number increases[10-12]. However, BP is extremely unstable under ambient conditions.The challenge of device stability has to be solved before it can be implemented for optoelectronics applications[13].

As a prototypical layered material in the III-VI family, InSe is a stable nonlinear optical crystal and exhibits superior electrical and optical properties[2,14-17]. Recently, the QHE has been demonstrated in a thin flake (6 layers) with mobility exceeding 10,000 cm$^2$/(V·s)[2]. Moreover, in contrast to the TMDCs, InSe is a direct bandgap material from the bulk to a few layers (> 5 layers), which renders it a promising candidate for efficient optoelectronic devices possessing large absorbance[15-17]. In addition, the bandgap of multilayer InSe is located in the infrared regime and bridges the spectrum gap between the TMDCs and BP[15].

One unique advantage of 2D materials is their ultrahigh flexibility[18]. It has been shown that a single-layer graphene can sustain as much as ~ 10% strain without mechanic failure[19], and similar findings have also been reported in other 2D semiconductors such as MoS$_2$[19-24]. Meanwhile, strain can effectively and reversibly change the band structure of 2D materials to achieve a tunable spectral response in optoelectronic devices[21-23,25-27]. Two-dimensional InSe flakes provide an ideal platform to realize such functionality in the infrared regime. Unlike the TMDCs, a few-layer InSe is a direct bandgap semiconductor[15,28]. Considering the relatively small exciton binding energy in a few-layer InSe flake[28], we expect that the optical bandgap obtained from the PL spectroscopy would correspond closely to the direct bandgap at the $\Gamma$ point for a few-layer (> 5) system in the bulk hexagonal BZ, which is folded from the $Z$ point in the BZ of rhombohedral primitive cell (Supporting Information). Therefore, PL spectroscopy can be employed to investigate the strain-induced bandgap modulation of few-layer InSe flake. The strain effect on the InSe bandgap, to our best knowledge, has not been explored experimentally.

In this work, we demonstrate that, for a typical few-layer InSe flake, we can reversibly modulate the bandgap by about 239 meV by spanning a uniaxial compressive strain up to 0.62% (increasing the bandgap by ~ 79 meV) and a uniaxial tensile strain up to 1.06% (decreasing the bandgap by ~ 160 meV). This substantial bandgap modification arises from a steep bandgap strain coefficient of ~ 140 meV/% for uniaxial compressive strain and ~ 154 meV/% for the uniaxial tensile strain. The observed bandgap strain coefficient is the most pronounced among what have been reported in 2D materials, which would not only make InSe-based optoelectronic devices highly tunable in spectral response but also would render them promising candidates for flexible and sensitive strain sensors. We also confirm the bandgap modification by photoconductivity spectroscopy in which we measure the photoconductivity as a function of excitation photon energy.

## 2. Methods

Experimentally, we exfoliated InSe flakes from the $\gamma$-InSe single crystal onto the flexible poly(ethylene terephthalate) (PET) substrate with a thickness of 120 μm. The strain was applied using a home-built instrument in which the PET film was clamped at both ends. By pushing the PET film from one clamped end through a micrometer manipulator, we can precisely control the lateral movement and determine either uniaxial tensile or compressive strain, as schematically shown in Fig. 1c. The quantitative strain value can be determined by a two-point bending method: The bent PET is assumed as a circular arc for the strain calculation with the equation: $\varepsilon = \tau / R$, where $2\tau$ and $R$ are the thickness of PET substrate and the radius of curvature of the bent PET, respectively (Fig. 1d)[20]. Details of the strain measurement method can be found in the Supporting Information (SI). We investigate the strain effect using Raman and PL spectroscopy in a confocal microscope setup. We also studied the strain effect using the photocurrent measurement. First-principle calculations using density function theory (DFT) and quasiparticle GW calculation are employed to understand our experimental results. The details of the calculation and crystal growth can be found in the Supporting Information.

## 3. Results and discussion

The atomic structure of rhombohedral $\gamma$-InSe is given in Fig. 1, consisting of In-Se-Se-In single layers stacked vertically with van der Waals interaction. Each layer is of a hexagonal structure with $D_{3h}$ symmetry. The layer-dependent bandgap of InSe has been investigated both theoretically and experimentally[15,28]. Bulk InSe is a direct bandgap semiconductor with both the conduction band minimum (CBM) and valence band maximum (VBM) located at the $Z$ point, forming a direct bandgap semiconductor. InSe flakes retain the direct bandgap feature until they are less than five layers thick[15,28].

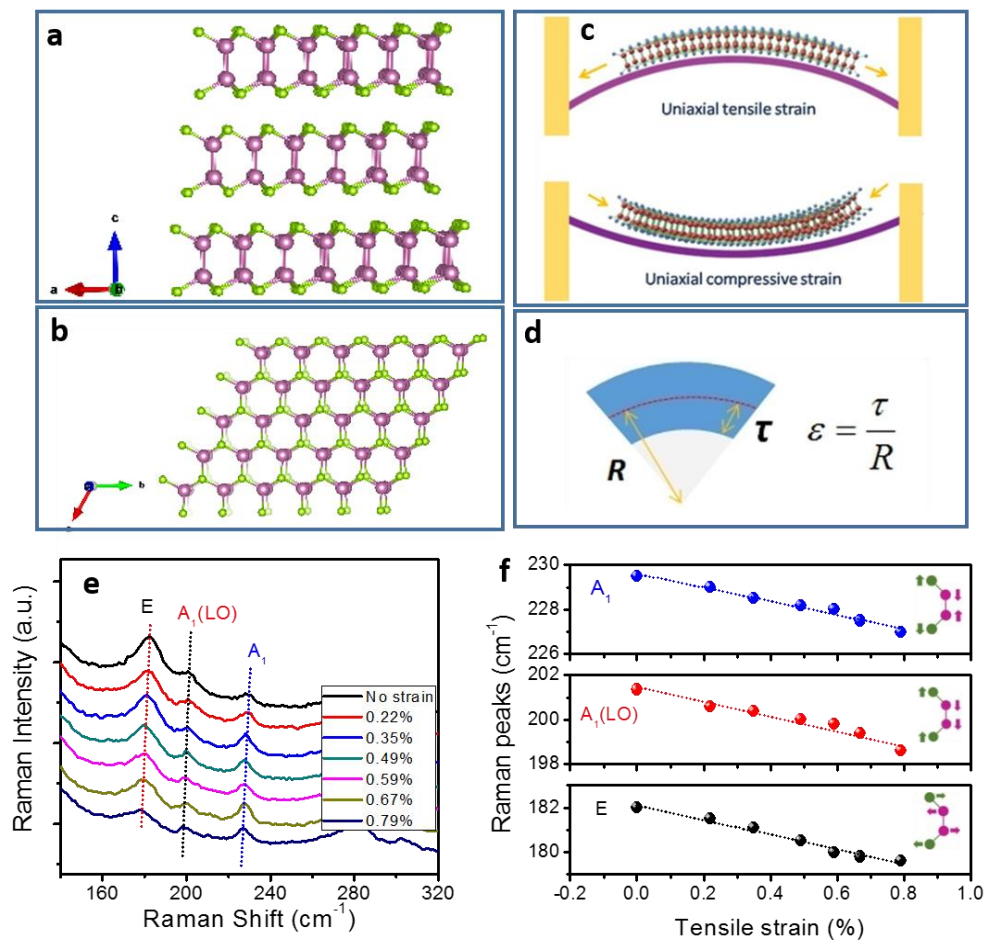

**Figure 1. Strain-induced Raman mode shifts in few-layer InSe.** (a) A side view of the atomic structure of the InSe crystal structure. In atom: purple. Se atom: green. (b) A top view of the InSe crystal structure showing a hexagonal structure with $D_{3h}$ symmetry. (c) A schematic of the two-point bending apparatus used for applying uniaxial tensile and compressive strain. (d) A schematic for the calculation of the strain in InSe flake. (e) Evolution of Raman spectra of an InSe flake of thickness ~ 8 nm with the tensile strain from 0 to 0.79%. The phonon vibration modes E, $A_1$(LO) and $A_1$ modes, are labeled. (f) The peak positions of E, $A_1$(LO) and $A_1$ Raman modes as a function of applied strain. Schematic representation of relative atomic motion of each Raman mode is also shown.

We first investigate the evolution of the Raman modes of InSe with tensile strain[20,21]. As shown in Fig. 1e (and Fig. S2 in the SI), the prominent peaks at 180, 198 and 226 cm$^{-1}$ for the unstrained InSe (~8 nm thick) on the PET film correspond to the doubly degenerate non-polar E mode (E'($\Gamma_3^1$) and E''($\Gamma_3^3$)), polar $A_1$(LO) mode and non-polar $A_1$ mode, respectively, consistent with previous reports[17,28]. Peak positions as a function of tensile strain are plotted and fitted to a linear function, shown in Fig. 1f, where all three modes exhibit linear shifts as a function of tensile strain at a rate of −3.3 (E mode), −3.37 ($A_1$(LO) mode) and −3.08 ($A_1$ mode) cm$^{-1}$/% strain, respectively. It is worth noting that we did not observe an obvious splitting of the doubly degenerate in-plane E mode at Raman shift of 180 cm$^{-1}$, which can be ascribed to the relatively small strain applied or to peak broadening[20,21]. Unlike the case of TMDCs in which only the in-plane Raman E mode is sensitive to the in-plane strain, both in-plane and out-of-plane Raman modes of InSe are shifted under strain, indicating that the tensile strain has significant effects on both vibrations along and perpendicular to the xy-plane. The change in the Raman E mode (~180 cm$^{-1}$) in few-layer InSe is mainly determined by the change of intralayer In-Se bonds[28], and a shift of the Raman E mode can be intuitively understood as the application of a tensile strain would increase the intralayer In-Se bond length. Meanwhile, the Raman $A_1$(LO) and $A_1$ modes are determined by the In-Se and In-In bond perpendicular to the xy-plane, respectively. Due to the Poisson contraction, these bond lengths and interlayer distance would change with strain, leading to the shift of the out-of-plane Raman $A_1$(LO) and $A_1$ modes[28]. The large shift of Raman $A_1$(LO) and $A_1$ modes in Fig. 1e and f also indicate that the effect of out-of-plane strain due to the Poisson contraction cannot be ignored in InSe flakes, different from that in the TMDCs.

The strain-dependent Raman E, $A_1$(LO) and $A_1$ modes enable us to calculate the parameters characterizing anharmonicity of the inter-atomic potential such as the Gruneisen parameter, which describes the effect that varying the volume of a crystal has on its vibrational properties. The Gruneisen parameter ($\gamma_m$) can be calculated by the following equation[33]:

$$\gamma_m = -\frac{\omega - \omega_0}{2\varepsilon(1-\nu)\omega_0},$$

where $\omega(\omega_0)$ is the frequency of a Raman mode with (without) strain, $\upsilon$ is Poisson's ratio (~ 0.4)[34], and $\varepsilon$ the tensile strain. The Gruneisen parameters ($\gamma_m$) of the E,

A$_1$(LO) and A$_1$ modes are determined to be ~ 1.51, ~ 1.39 and ~ 1.12 for the few-layer InSe shown in Fig. 1e, respectively, consistent with previous experimental results induced from the temperature-dependent Raman characterization of bulk InSe[35].

We further investigate the evolution of the InSe bandstructure with strain using photoluminescence (PL) spectroscopy and first-principles calculations. Drastically different from semiconducting TMDCs, our first-principles calculation shows that the unstrained InSe with thickness more than ~ 4 nm (about five-layer with a single layer thickness of 0.82 nm) can be approximately treated as a direct-bandgap semiconductor, with $\left|E_g^\Gamma - E_g^{ind}\right| < 30$ meV based on DFT calculations, where $E_g^\Gamma$ is the direct bandgap at $\Gamma$ point, and $E_g^{ind}$ the indirect bandgap (SI). Consequently, with the neglect of possible excitonic effects, the PL spectroscopy directly reads out the optical bandgap of InSe through the PL peak position. Considering the significant dielectric screening when the layer number exceeds five (exciton binding energy will be small)[11], the optical bandgap of InSe is approximately the same as the quasi-particle bandgap. Fig. 2a shows the PL spectra of multilayer InSe of thickness ~ 12 nm over a range of 0–1.06% tensile strain. The main PL peak in the unstrained InSe is located at 1.34 eV. With increasing strain, the PL spectra of InSe are significantly changed: the peak exhibits a redshift of about 118 meV with a tensile strain of ~1.06%. We plot the PL peaks as a function of strain value for this sample in Fig. 2b (blue diamonds) with a linear fitting. The slope, demonstrating the sensitivity of the optical bandgap to strain, is ~ 101 meV/% strain for this sample. We also explored the effect of strain on the bandgap of InSe flakes of different thicknesses. For all the flakes with thickness ranging from 4 nm to more than 30 nm, the PL peak exhibits a consistent red shift with increasing tensile strain, as shown in Fig. 2b, and the peak shift rate is determined to be ~154±8 meV/% for a thin flake (4 and 6 nm) and 81±4 meV/% strain for a thick flake (> 30 nm).

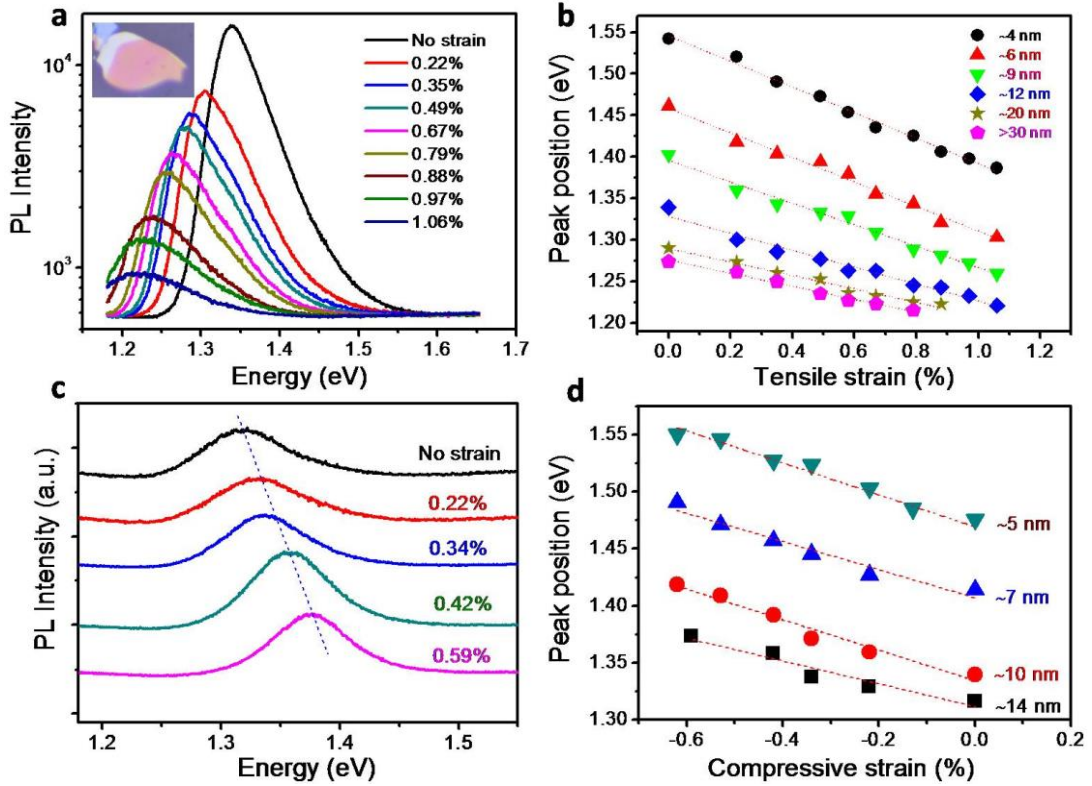

**Figure 2. Strain-induced bandgap change revealed by PL spectroscopy.** (a) PL spectra of a multilayer InSe of thickness ~ 12 nm. The applied uniaxial tensile strain ranges from 0 to 1.06%. Inset: the optical image of the flake on PET. The laser excitation spot is at the center of the pink flake. (b) PL peak positions versus tensile strain in InSe flakes of different thicknesses. (c) PL spectra of a multilayer InSe flake of thickness ~14 nm under different uniaxial compressive strains. (d) PL peak position versus compressive strain in InSe flakes of different thicknesses.

We also investigate the effect of the compressive strain on the bandgap by applying a relatively small strain (from 0 to 0.62%), with the purpose of avoiding complication due to potential buckling or other mechanical deformation of the flakes. As shown in Fig. 2c, the PL peak of InSe of thickness ~14 nm blueshifts by about 56 meV when the compressive strain approaches 0.59%, corresponding to a shift rate of ~ 100 meV/% strain. These results show that the bandgap of InSe can be increased through compressive strain, compared with a decrease of the bandgap achieved with tensile strain. We also measured the compressive strain dependent PL of samples with different thicknesses, and the PL peak position as a function of strain is shown in Fig. 2d. The bandgap of InSe flakes obtained from the PL spectra increases with strain, and the bandgap shift rate for a few-layer InSe flake (~ 5 nm) is determined to be ~

140 meV/% strain. The combination of tensile strain and compressive strain greatly enhances the bandgap tunability of an InSe flake.

**Table 1.** Comparison of bandgap tunability between InSe and other one- or two-dimensional semiconductors under uniaxial tensile, compressive and biaxial strain. The InSe data with tensile strain (positive value) and compressive strain (negative value) are highlighted in the left column with orange and light yellow background, respectively. Previously reported data from other 1D or 2D materialsare organized in the right column (highlighted in blue) for comparison.

| Materials | Thickness (nm) | Strain (%) | Strain coefficient (meV/%) | Materials | Thickness (nm) | Strain (%) | Strain coefficient (meV/%) |
|---|---|---|---|---|---|---|---|
| InSe | 4 | 1.06 | 153 | [21,22]$MoS_2$ | Monolayer | 2.2 | 45 |
| | 6 | 1.06 | 154 | | Monolayer | 0.52 | 70 |
| | 9 | 1.06 | 132 | [23,43]$WSe_2$ | Monolayer | 1.04 | 60-70 |
| | 12 | 1.06 | 101 | | Monolayer | 1.4 | 54 |
| | 20 | 0.88 | 88 | [36]$MoSe_2$ | Monolayer | 1.1 | 27 |
| | 35 | 0.79 | 81 | [37]$WS_2$ | Monolayer | 2.5 | 20-30 |
| | | | | [40]$ReSe_2$ | Monolayer | 1.64 | 36 |
| InSe | 5 | −0.62 | 140 | [38,39]B-P | 10 | 5 | 100-140 |
| | 7 | −0.62 | 123 | | 4 | 0.92 | 117-124 |
| | 10 | −0.62 | 133 | [41]$MoS_2$ Biaxial strain | Monolayer | >2 | 99 |
| | 14 | −0.59 | 100 | [42]GaAs | Nanowire 40 nm | 3.5 | 84 |

We compare our results with previously reported values in other one- and two-dimensional materials, which we summarized in Table 1. It is evident that the bandgap strain coefficient with a uniaxial tensile strain in few-layer InSe is the most pronounced among what have been reported experimentally in 2D materials so far, several times larger than those of TMDCs and comparable to that of BP [21-23,25,36-39]. This excellent tunability may finds its applications in the future development of flexible electronic, optoelectronic and photonic deviceswith near-infraredspectral response.

To understand the microscopic mechanism of the strain-induced bandgap change, we performed first-principles density functional theory (DFT) calculations and quasi-particle GW calculations on InSe at the single-layer (Fig. 3a) and the bulk limit (Fig. 3b). We did not include spin-orbit coupling (SOC) in the shown calculations because SOC has a minimum effect on the bandgap studied here (see Supporting Information), considering the conduction band minimum (CBM) is mainly composed of In *s* orbitals and the valence band maximum (VBM) is mainly composed of Se $p_z$ orbitals[17,28].

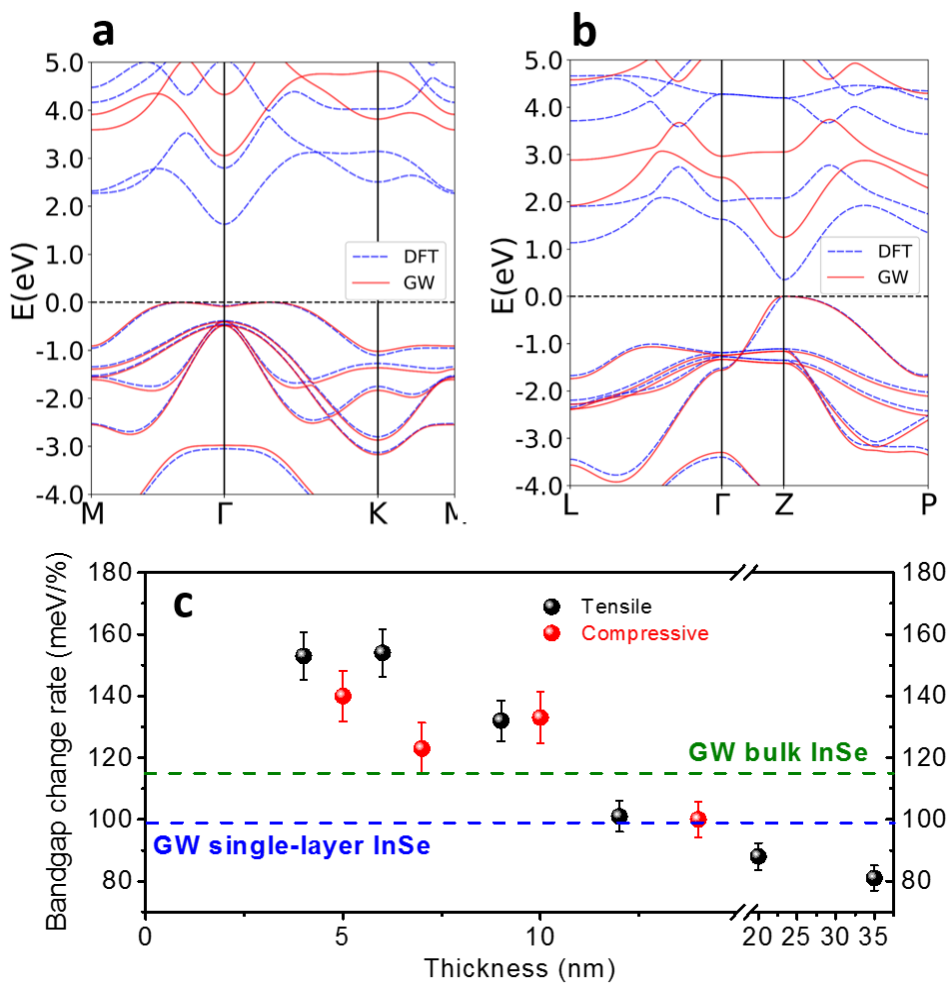

**Figure 3. Calculation of the strain induced bandgap change in InSe.** (a) DFT-PBEsol and GW band structure of single-layer InSe. (b) DFT-PBEsol and GW band structure of bulk InSe. (c) Experimentally measured bandgap strain coefficients for uniaxial tensile strain (black dots) and compressive strain (red dots) as a function of the InSe flake thickness. The blue and green dashed lines label the calculated quasi-particle bandgap strain coefficient of single layer and bulk InSe, respectively.

Unstrained single-layer InSe is found to be an indirect bandgap semiconductor, with a "Mexican hat" feature near the $\Gamma$ point. As shown in Fig. 3a, the direct bandgap at $\Gamma$ is ~1.7 eV from DFT calculations, while GW calculations show a bandgap of ~ 3.1 eV.

This drastic difference in the bandgap between the two calculations originates from strong electron-electron interaction and weak dielectric screening in quasi-2D materials. As the layer number increases to more than five layers, DFT calculations show that InSe quickly becomes a direct bandgap semiconductor, with the bandgap located at the $\Gamma$ point (See Supporting Information). The band structure of unstrained bulk InSe is shown in Fig. 3b. Bulk InSe is found to be a direct bandgap semiconductor with the bandgap located at the $Z$ point in the Brillouin zone (BZ) of the rhombohedral primitive cell. The bandgap is 0.35 eV from DFT calculations and 1.25 eV from GW calculations. As for other semiconductors, the GW bandgap value shows an excellent agreement with the experimentally measured bandgap which is at ~ 1.27 eV for a 30 nm thick InSe layers.

Since we access the direct bandgap experimentally through PL spectroscopy, in our calculations, we focus on the strain effect on the corresponding direct bandgap for the single-layer and bulk InSe. We first investigate the strain effect on the single-layer InSe, and our DFT calculation shows that the bandgap strain coefficient is 91 meV/% for an in-plane uniaxial tensile strain (see Supporting Information). Our GW calculation shows a coefficient of 97 meV/% (Fig. 3c), only 7% larger than that from the DFT calculation. We also calculated the bandgap strain coefficient for bulk InSe, and the DFT calculation shows a value of 95 meV/% and the GW calculation results in an increased value of 114 meV/% (Fig. 3c).

We plot the experimental bandgap strain coefficients for InSe of different thicknesses in Fig. 3c and compare them with the GW calculation results. Since PL peaks can be determined accurately in experiments, the uncertainty in measurement is mainly from the strain calculation (see Supporting Information). Fig. 3c shows that for InSe flake of thickness ~ 4 nm (5 layers) and ~ 6 nm (8 layers), the bandgap strain coefficient due to uniaxial tensile strain is almost the same, within the experimental uncertainty considered. The bandgap strain coefficient due to uniaxial compressive strain, which is slightly smaller, is also about the same for the InSe flake of thickness from ~ 5 nm to ~ 10 nm. The value of the nearly constant bandgap strain coefficient is in better agreement with the GW calculation for bulk InSe (114 meV/%), and larger than what is predicted for the single-layer InSe (97 meV/%). In addition, we attribute the remaining discrepancy between experimental and calculated bandgap strain coefficients to unintended out-of-plane strain in the experiment, which is confirmed by the strain-induced shift of Raman $A_1$ and $A_1$(LO) modes in Fig. 1e. Calculations with additional out-of-plane tensile (compressive) strain for bulk InSe structures with fixed in-plane compressive (tensile) strain will increase the calculated bandgap strain coefficient (Sec. 10 in SI). It is interesting to note that the strain-induced effect on the bandgap in few-layer InSe is closer to that of bulk InSe rather than a single-layer InSe. This can be intuitively understood by the fact that the electronic bandstructure of few-layer InSe is closer to that of the bulk InSe (direct bandgap at the $\Gamma$ point: ~1.3 eV) shown in Fig. 3b, much smaller than that of the single-layer InSe at the $\Gamma$ point (3.14eV) shown in Fig. 3a.

We also note that InSe flakes thicker than 10 nm show significantly decreased bandgap strain coefficient as the thickness of the flake increases. This trend is likely due to an effective strain decay in the thick flakes. The nominal strain is calculated based on the geometry of the flexible PET film, which should be the same as that in the thin InSe flake since it exactly follows the change of the substrate[20,21]. For thick samples, we expect inefficient strain transfer from PET film to InSe flake or inefficient strain transfer between different layers. We believe that the latter plays an important role and it is confirmed by the extra broadening observed in PL width of strained thick InSe flake (see Supporting Information). Both effects will reduce the effective strain applied to the flake and lead to reduced strain coefficient observed. The bandgap strain coefficient for bulk InSe is calculated as ~ 97 meV/% for DFT and ~ 114 meV/% for GW, both larger than the observed value (~ 81 meV/%) from thick InSe of ~ 35 nm thickness.

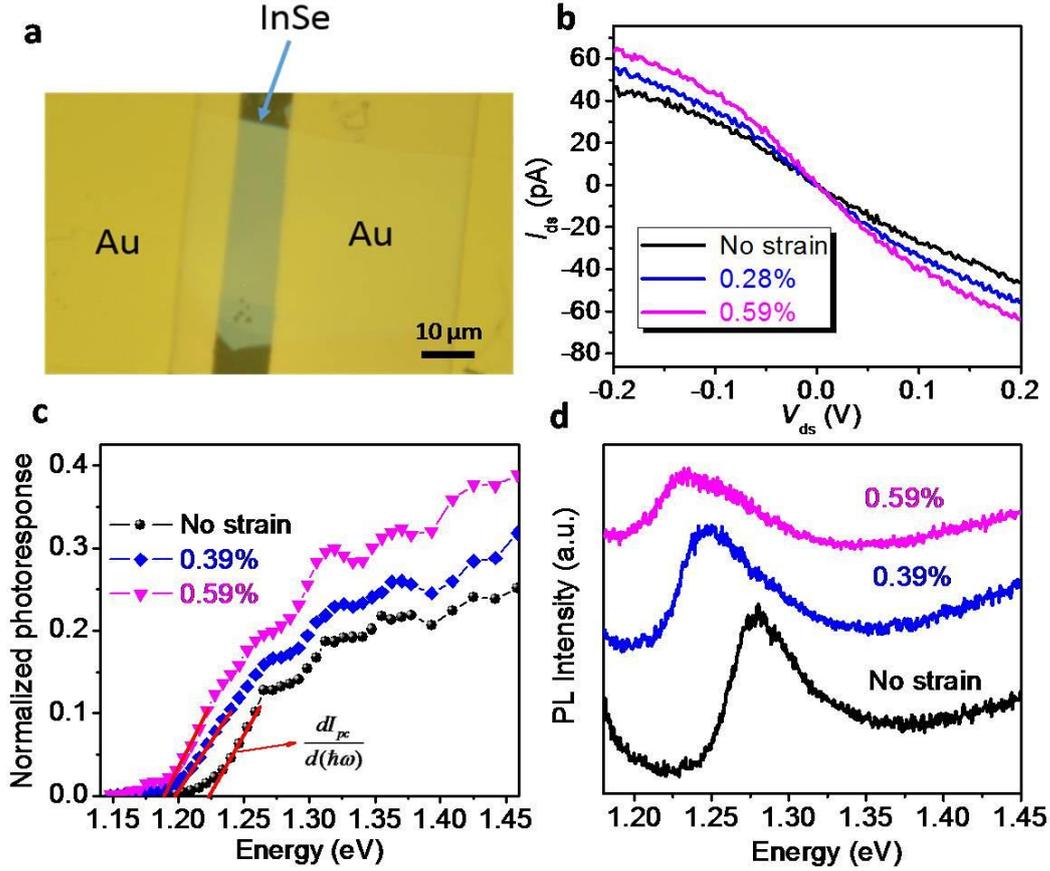

**Figure 4. Piezoresistivity and photoconductivity of the strained InSe flake.** (a) Optical image of an InSe device on the PET substrate. (b) *I-V* curves of InSe of thickness ~ 35 nm under uniaxial tensile strain. (c) Normalized photocurrent of unstrained and tensile-strained InSe as a function of excitation photon energy. (d) The corresponding PL spectra of InSe under different tensile strains.

The controllable modification of InSe bandgap enables the modulation of resistivity of InSe based devices. To demonstrate a relevant application in photodetection, we investigated the photocurrent response of an InSe flake of thickness ~35 nm subjected to a tensile strain. The device was fabricated by exfoliation of an InSe flake directly onto a PET substrate, followed by e-beam evaporation of 5/50 nm Ti/Au to define electrodes using a shadow mask, avoiding possible contamination induced by photoresist or e-beam resist residue. The optical image of the device is shown in Fig. 4a. According to Fig. 4b, the conductance increases by 1.5 times at a bias of 200 mV as the tensile strain is increased to 0.59%. Due to the non-centrosymmetric character of multilayer γ-InSe flakes, there are two effects potentially contributing to the change of transport behavior: the piezoelectric effect, in which strain-induced charge asymmetrically modulates the Schottky barriers at the electrical contacts; and the piezoresistive effect, in which the strain-induced bandgap change modulates the resistance of the device. The nearly symmetric modulation of *I-V* curves in Fig. 4b indicates that the piezoresistive effect in InSe dominates the change of electrical transport behavior, which can be explained with the reduced bandgap with a tensile strain applied[44]. The sensitive conductance dependence on strain can be qualitatively described by the gauge factor (*g*) of InSe, whichis defined as $g=(I-I_0)/(\varepsilon I_0)$, where $I$ and $I_0$ are the currents with and without strain, respectively[44]. The gauge factor is estimated to be ~84 for the device shown in Fig. 4a, which is comparable to that of the state-of-the-art silicon strain sensors[45].

We further measured the photoconductivity response to excitation light with different wavelength. The excitation-wavelength-dependent photoconductivity measurement has been applied previously to measure the band edge of semiconductors with small exciton binding energy[46,47]. For this particular measurement, we intentionally broadened the laser beam spot to ~6 mm to illuminate the entire device uniformly. We used a broadband laser source (super-continuum white laser) and selected the excitation wavelength with a tunable filter (width ~ 6 nm). To avoid nonlinear response or possible damage, a low excitation power of 30 to 120 μW was used. By applying a DC bias voltage of 300 mV, the photocurrent was measured through a lock-in amplifier by modulating the excitation laser through a mechanical chopper. The normalized photoresponse ($I_{pc}/P_{power}$) as a function of wavelength is shown in Fig. 4c. In the absence of strain, we observe an abrupt increase of the photocurrent when the excitation photon energy exceeds 1.2 eV. We estimate the band edge by taking derivative of current vs. excitation energy ($dI_{pc}/d(\hbar\omega)$), the red line in Fig. 4c)[47]. Fig. 4c suggests that the bandgap of the unstrained InSe flake is ~ 1.22 eV, consistent with the value extracted from PL measurement (~ 1.27eV) shown in Fig. 4d. As we increase the tensile strain to 0.39%, the onset energy of photocurrent is redshifted to 1.19 eV. The bandgap shift (~ 27 meV) extracted from the photocurrent measurement is consistent with the shift (~ 35 meV) from PL measurement shown in Fig. 4d. When the strain is increased from 0.39% to 0.59%, the PL peak shows a further redshift of ~ 11 meV, in agreement with the photocurrent measurement (7±2 meV). These observations confirm that the uniaxial tensile strain indeed decreases the bandgap of the multilayer InSe and thus broadens the absorption spectrum range. This sensitive modulation of photocurrent response from InSe, through the modulation of the direct bandgap, provides a new knob to engineer and control novel optoelectronic devices based on InSe.

## 4. Conclusions

In summary, we have investigated the effect of uniaxial tensile and compressive strain on the electronic structure of InSe with different thicknesses. The strain-induced phonon-softening and bandgap modulation were experimentally demonstrated, suggesting that strain engineering is an effective tool to explore the novel physics of III-VI layered compounds. Applying a relatively small strain, we have demonstrated a reversible optical bandgap tunability of ~239 meV in the infrared regime. Such tunability originates from a sensitive response of InSe electronic structure to the external strain, which manifests itself as a substantial bandgap strain coefficient as large as ~ 154 meV/% under a uniaxial tensile strain. The observed bandgap strain coefficients are confirmed by our first-principles calculations. This sensitive response to strain holds great potential for a wide range of applications based on atomically thin InSe flakes, including electromechanical, piezoelectric and optoelectronic devices.

**Supporting information**

Estimation of strain uncertainty, details of Raman and PL spectra of InSe flakes, PL spectra of InSe wrinkles, DFT and GW calculation results of InSe, crystal growth of InSe.

**Acknowledgements**

The device fabrication was supported by Micro and Nanofabrication Clean Room (MNCR), operated by the Center for Materials, Devices, and Integrated Systems (cMDIS) at Rensselaer Polytechnic Institute (RPI). Tianmeng Wang and Yanwen Chen acknowledge the support of the Howard Isermann Fellowship from the Department of Chemical and Biological Engineering of Rensselaer Polytechnic Institute. Sufei Shi acknowledges the startup fund from Rensselaer Polytechnic Institute. The optical characterization was supported by the Center for Future Energy Systems (CFES), a New York State Center for Advanced Technology at RPI. Cheng-Yan Xu acknowledges the support from National Natural Science Foundation of China (No. 51572057). The theoretical studies were supported by Theory of Materials Program at the Lawrence Berkeley National Laboratory through the Office of Basic Energy Sciences, US Department of Energy under Contract No. DE-AC02-05CH11231, which provided for the first-principles GW-BSE calculations, and by the National Science Foundation under Grant No. DMR-1508412, which provided for the DFT calculations. Computational resources were provided by the DOE at Lawrence Berkeley National Laboratory's NERSC facility and the NSF through XSEDE resources at NICS.